# Experimental study of Hong-Ou-Mandel interference using independent phase randomized weak coherent states

Eleftherios Moschandreou, Jeffrey I. Garcia, Brian J. Rollick, Bing Qi, Raphael Pooser, and George Siopsis

*Abstract*—**Hong-Ou-Mandel interferometers are valuable tools in many Quantum Information and Quantum Optics applications that require photon indistinguishability. The theoretical limit for the Hong-Ou-Mandel visibility is 0.5 for indistinguishable weak coherent photon states, but several device imperfections may hinder achieving this value experimentally. In this work, we examine the dependence of the interference visibility on various factors, including (i) detector side imperfections due to after-pulses, (ii) mismatches in the intensities and states of polarization of the input signals, and (iii) the overall intensity of the input signals. We model all imperfections and show that theoretical modeling is in good agreement with experimental results.**

*Index Terms*—**Quantum key distribution, interference, avalanche photodiodes, coherent state, visibility, Bell state.**

## I. INTRODUCTION

THE interference of two photons at a beam splitter was first examined by Hong, Ou, and Mandel [1] (HOM interference). As the input photons (Fig. 1) become increasingly indistinguishable in all degrees of freedom, the coincidence rate of the beam-splitter-output photons exhibits a characteristic dip, the depth of which depends on the degree of indistinguishability of the input photons [2]. Various applications have been proposed and demonstrated, utilizing the interference of single photons created through Spontaneous Parametric Down Conversion (SPDC), including clock synchronization [3], quantum teleportation [4], and quantum logic gates [5].

A convenient alternative to SPDC heralded photons is an input state consisting of weak coherent states [6], implemented as attenuated laser light. Studies have been conducted to examine the HOM visibility using coherent states, including the effect of the laser frequency chirp and time jitter [7], [8], the optical delay between the inputs and detection time differences [9], and frequency mismatch [10].

Because HOM interference can be used for experimental Bell state analysis [11], [12], it lies at the heart of Measurement-Device-Independent (MDI) Quantum Key Distribution (QKD)

[12], which is a novel quantum communication protocol resistant in all possible detector-side attacks. The applicability of the protocol has been demonstrated in multiple experiments [14]-[19]. In MDI QKD, the interference visibility significantly affects the final key generation rate [7], [19], [20]. The use of coherent states instead of single-photon states could open a potential vulnerability due to the non-zero probability of multiple photon pulses, but the implementation of the decoy-state method [21]-[23] can overcome such a threat.

Wang, *et al.*, [24] examined how realistic imperfections of the devices used in an HOM interference experiment affect the HOM visibility. In particular, they considered possible imperfections of the beam splitter, mismatches in the input intensities, and studied the effect of the after-pulses in single-photon avalanche detectors. In this work, we provide experimental measurements and extend the work of [24] to include possible mismatches in the state of polarization of the inputs and examine the effect of the overall intensity of the inputs on the HOM visibility. We also discuss modeling of imperfections and show good agreement of experimental results with the theoretical modeling.

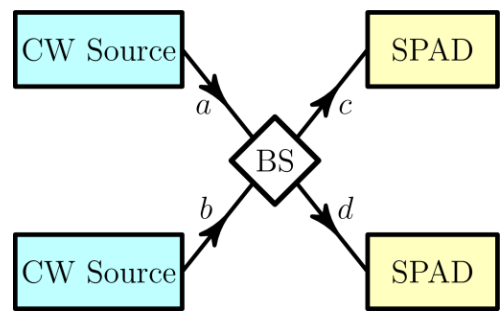

Fig. 1. Schematic of the experimental set-up. Two weak coherent pulses enter the *a* and *b* ports of the beam splitter (BS) and interfere. Each output port (*c* and *d*) is directed to a single-photon avalanche InGaAs detector (SPAD).

This work was supported by the U.S. Office of Naval Research under award number N00014-15-1-2646.

The authors are with the Department of Physics and Astronomy, University of Tennessee, Knoxville, Tennessee 37996-1200, U.S.A. (e-mail:

emoschan@vols.utk.edu; jgarcia1@vols.utk.edu; brollick@vols.utk.edu; qib1@ornl.gov; pooserrc@ornl.gov; siopsis@tennessee.edu).

B. Qi and R. Pooser are also with the Quantum Information Science Group, Computational Sciences and Engineering Division, Oak Ridge National Laboratory, Oak Ridge, Tennessee 37831-6418, U.S.A.

## II. PARAMETRIZING THE HONG-OU-MANDEL INTERFERENCE VISIBILITY

The set-up for our HOM interference measurements consists of two independent input laser pulses, interfering at a beam splitter (BS) and with each output directed to a single-photon avalanche detector (SPAD) (Fig. 1).

We model the input to the beam-splitter state as two weak coherent states:

$$| \Psi_{in} \rangle = | \alpha \rangle \otimes | \beta \rangle = e^{-\frac{\mu_a + \mu_b}{2}} e^{\alpha \hat{a}^\dagger + \beta \hat{b}^\dagger} | 0 \rangle , \qquad (1)$$

created by creation operators $\hat{a}^\dagger$ and $\hat{b}^\dagger$, and of parameters $\alpha$ and $\beta$, respectively. The coherent-state parameters are complex and include a phase, and $\mu_{a,b}$ are the corresponding average photon numbers of the two beams ($\mu_a = |\alpha|^2$ and $\mu_b = |\beta|^2$). In our experimental setup, the phases are randomized. Therefore, the initial state is

$$\rho_{in} = \int_0^{2\pi} \frac{d\theta_a}{2\pi} \int_0^{2\pi} \frac{d\theta_b}{2\pi} | \Psi_{in} \rangle \langle \Psi_{in} |$$
$$= e^{-\mu_a - \mu_b} \sum_{m,n=0}^{\infty} \frac{\mu_a^m \mu_b^n}{(m!n!)^2} \left( \hat{a}^\dagger \right)^m \left( \hat{b}^\dagger \right)^n | 0 \rangle \langle 0 | \hat{a}^m \hat{b}^n . \qquad (2)$$

Nevertheless, we will continue to work with the state (1) and average over the phases at the end. To account for the action of the beam splitter, we introduce a pair of orthogonal directions, named horizontal and vertical, respectively, and express the polarization vectors of the incoming beams $\hat{\varepsilon}_{a,b}$ in terms of unit vectors in the chosen directions, $\hat{\varepsilon}_{H,V}$. The creation operators are similarly expressed as linear combinations:

$$\hat{a}^\dagger = \hat{\varepsilon}_a \cdot \hat{\varepsilon}_H \, a_H^\dagger + \hat{\varepsilon}_a \cdot \hat{\varepsilon}_V \, a_V^\dagger ,$$
$$\hat{b}^\dagger = \hat{\varepsilon}_b \cdot \hat{\varepsilon}_H \, b_H^\dagger + \hat{\varepsilon}_b \cdot \hat{\varepsilon}_V \, b_V^\dagger . \qquad (3)$$

The action of a beam splitter with reflectivity $R=r^2$ and transmissivity $T=t^2$, with $R+T=1$, is described by the unitary transformation:

$$\hat{a}_i^\dagger \to t\hat{c}_i^\dagger + r\hat{d}_i^\dagger ,$$
$$\hat{b}_i^\dagger \to r\hat{c}_i^\dagger - t\hat{d}_i^\dagger , \qquad (4)$$

where $c_i^\dagger$ and $d_i^\dagger$ are the creation operators of the respective output beams, with $i = H, V$. The input state (1) transforms into the output state

$$| \Psi_{out} \rangle = e^{-\frac{\mu_a + \mu_b}{2}} \prod_{i=H,V} e^{\alpha \left( t\hat{c}_i^\dagger + r\hat{d}_i^\dagger \right) \hat{\varepsilon}_a \cdot \hat{\varepsilon}_i} e^{\beta \left( r\hat{c}_i^\dagger - t\hat{d}_i^\dagger \right) \hat{\varepsilon}_b \cdot \hat{\varepsilon}_i} . \qquad (5)$$

Given this output state, the probability $P_{mn}$ that $m$ $(n)$ photons emerge at output port $c$ $(d)$ is found to be (see Appendix for details)

$$P_{mn}^{(out)} = e^{-\mu_c - \mu_d} \frac{\mu_c^m \mu_d^n}{m!n!} , \qquad (6)$$

where $\mu_{c,d}$ are the corresponding mean photon numbers at the two output ports of the beam-splitter,

$$\mu_c = \mu_a t^2 + \mu_b r^2 + 2tr \Re \left( \alpha \beta^* \hat{\varepsilon}_a \cdot \hat{\varepsilon}_b^* \right) ,$$
$$\mu_d = \mu_a r^2 + \mu_b t^2 - 2tr \Re \left( \alpha \beta^* \hat{\varepsilon}_a \cdot \hat{\varepsilon}_b^* \right) . \qquad (7)$$

Notice that the mean photon numbers of the beams obey the conservation law

$$\mu_a + \mu_b = \mu_c + \mu_d \qquad (8)$$

which is a consequence of the unitarity of the beam-splitter transformation (4), $R+T = r^2 + t^2 = 1$.

Our real detectors at the two beam-splitter ports have efficiencies $\eta_c$ and $\eta_d$, and dark-count probabilities $d_c$ and $d_d$, respectively. Therefore, the probability that the detectors click is given by

$$P_{mn} = P_{mn}^{(out)} \left( 1 - (1-\eta_C)^m (1-d_c) \right)$$
$$\times \left( 1 - (1-\eta_D)^n (1-d_d) \right). \qquad (9)$$

The total coincidence probability is given by

$$P^{(coin)} = \sum_{m,n=0}^{\infty} P_{mn} . \qquad (10)$$

After averaging over the phases, we obtain the total coincidence probability corresponding to the state (2) (see Appendix for details) in terms of Bessel functions:

$$P^{(coin)} = 1 - \mathcal{C} I_0 \left( 2\eta_c \sqrt{\mu_a \mu_b} \, tr \cos \Phi \right)$$
$$- \mathcal{D} I_0 \left( 2\eta_d \sqrt{\mu_a \mu_b} \, tr \cos \Phi \right)$$
$$+ \mathcal{C} \mathcal{D} I_0 \left( 2(\eta_c - \eta_d) \sqrt{\mu_a \mu_b} \, tr \cos \Phi \right), \qquad (11)$$

where

$$\mathcal{C} = e^{-\eta_c (\mu_a t^2 + \mu_b r^2)} (1 - d_c) ,$$
$$\mathcal{D} = e^{-\eta_d (\mu_a t^2 + \mu_b r^2)} (1 - d_d) , \qquad (12)$$

and $\Phi$ is a measure of the polarization mismatch between the two incoming beams defined by

$$\cos \Phi = \left| \hat{\varepsilon}_a \cdot \hat{\varepsilon}_b^* \right| . \qquad (13)$$

The total probability that the detector at port $c$ clicks, after averaging over phases, is also expressed similarly in terms of a Bessel function:

$$P^{(c)} = 1 - \mathcal{C} I_0 (2\eta_c \sqrt{\mu_a \mu_b} \, tr \cos \Phi). \qquad (14)$$

The total probability that the detector at port $d$ clicks is found similarly:

$$P^{(d)} = 1 - \mathcal{D} I_0 (2\eta_d \sqrt{\mu_a \mu_b} \, tr \cos \Phi). \qquad (15)$$

Details can be found in Appendix.

We define the Hong-Ou-Mandel visibility by

$$V_{HOM} = 1 - \frac{P^{(coin)}}{P^{(c)} P^{(d)}} . \qquad (16)$$

Using the explicit expressions (11), (14), and (15), we find that $0 \leq V_{HOM} \leq 0.5$. We aim at maximizing the value of $V_{HOM}$.

## III. EXPERIMENTAL SETUP

Our experimental setup is shown on Fig. 2. Two independent continuous-wave (CW) lasers (Wavelength References) at 1550 nm were employed to prepare weak coherent states. The frequency difference between the two lasers stayed below 10 MHz without performing any feedback control. Note, in all experiments, the phase difference between the two lasers swept through a multi-$2\pi$ range within the data acquisition time. This is equivalent to the phase averaging process assumed in the theoretical analysis. To generate laser pulses, two LiNbO$_3$ (EOSPACE) intensity modulators were used to modulate the outputs of the two lasers. The two intensity modulators were driven by the same digital delay generator (Stanford Research Systems) and their DC bias voltages were carefully adjusted to achieve high extinction ratios. The polarization state of each pulse can be changed with a homemade high-speed polarization modulator, which is driven by a Keysight Waveform Generator (WG). Details about the polarization modulator can be found in [16]. It is imperative that the WGs controlling the polarizations of Alice and Bob and the SRS share the same time base. The pulses enter an additional polarization controller where the state of polarization of each arm is fine-tuned, while the detection coincidences are monitored in real time, to achieve optimal HOM Visibility prior to data collection. The pulse is then digitally attenuated in order to reach the single-photon level. The pulses are lead to travel a free-space path before they interfere at the beam splitter and be read by single-photon avalanche detectors (SPAD). It should be noted that the free-space path is not needed in this experiment, but was introduced because of our interest in building a free-space QKD system. In future work, we will discuss additional imperfections due to effects specific to free space, such as turbulence.

The detectors are both IdQuantique 210 with one being an ultra-low noise model and operate in gated mode triggered by the SRS delay generator. Timestamps of the open gates and the detection events are recorded on a time-interval analyzer (TIA) with a resolution of 81 ps. The timestamps are finally analyzed to extract coinciding detection events and coinciding open gates. In our measurements the HOM probability is given by the ratio of the coincidence detection events over all the coinciding gates.

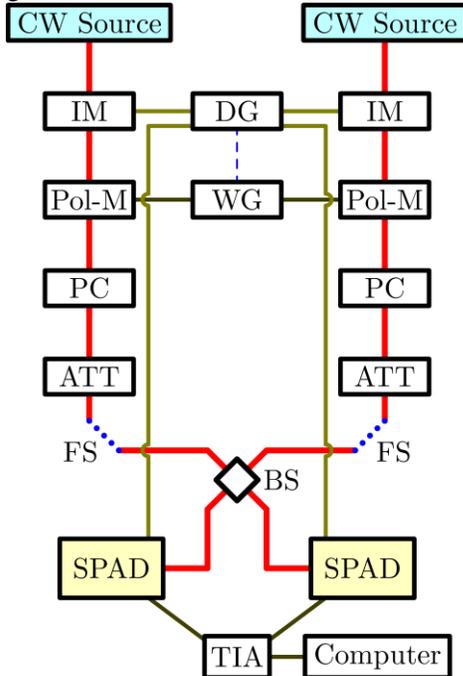

Fig. 2. Schematic of our experimental setup. Components: CW laser source, Intensity Modulator (IM) driven by a delay generator(DG), Polarization Modulator (Pol-M) driven by a waveform generator (WG), Manual Polarization Controller (PC), Single-Photon Avalanche Detector (SPAD) triggered by the delay generator, Digital Attenuator (ATT), Free space path (FS), 50:50 Beam-Splitter (BS).

## IV. RESULTS

Here we report on our experimental results. We examine how the HOM Visibility is affected by the after-pulse effect and by various imperfections in the source preparation.

### A. HOM Visibility and Detector Imperfections

We consider the effect on HOM visibility due to detector imperfections. Ref. [24] highlights the after-pulse effect as a significant source of error in an experimental implementation of the HOM interference. The authors of [24] employed a non-Markovian model and showed that the coincidence probability, after considering the after-pulse effect, can be written as:

$$P^{(coin;aft)} = P^{(coin)} + \left[P^{(c)} - P^{(coin)}\right]P^{(d)}P_d^{(total;aft)} + \left[P^{(d)} - P^{(coin)}\right]P^{(c)}P_c^{(total;aft)} \quad (17)$$

where $P^{(coin)}$ is the coincidence probability given by (11), and $P^{(c)}$, $P^{(d)}$ are the detection probabilities for the detectors at ports $c$ and $d$, respectively, given by (14) and (15). In (17), $P_c^{(total;aft)}$ and $P_d^{(total;aft)}$, describe the total after-pulse probability for each detector. We assume that the after-pulse probability decays with time as a simple exponential $P_{(t)} = P_0 \cdot e^{-t/\tau}$, with $P_0$ the initial after-pulse probability and $\tau$ the characteristic decay time. In gated mode the total after-pulse probability, receives contributions only when the gate is open:

$$P^{(total;aft)} = P_{(T_{dt})} + P_{(T_{dt}+T_{gat})} + \cdots = P_0 \frac{e^{-T_{dt}/\tau}}{1-e^{-T_{gat}/\tau}} \quad (18)$$

with $T_{dt}$ the detector dead time and $T_{gat}$ the gating period.

Probabilities $P^{(c)}$ and $P^{(d)}$ of (14) and (15) are similarly modified as,

$$P^{(c,aft)} = P^{(c)}\left[1 + (1-P^{(c)})P_c^{(total;aft)}\right] \quad (19)$$

$$P^{(d,aft)} = P^{(d)}\left[1 + (1-P^{(d)})P_d^{(total;aft)}\right] \quad (20)$$

With (18) we can relate the HOM visibility with the dead time settings on our detectors. First, we need to determine the parameters $P_0$ and $\tau$ experimentally. We follow the procedure described in [27] and collect histograms of detection events binned into time intervals between successive detection events. By fitting the logarithm of the frequencies with equation (6) in [27] we extract the parameters $P_0$ and $\tau$.

For our first run the gating and pulse frequency was set to 2MHz. The detection histograms presented in Fig. 3 gave the values $P^C{}_0$ =0.018 and $\tau^C$ =0.85μs for detector C, and $P^D{}_0$ = 0.033 and $\tau^D$ = 1.41 μs for detector D.

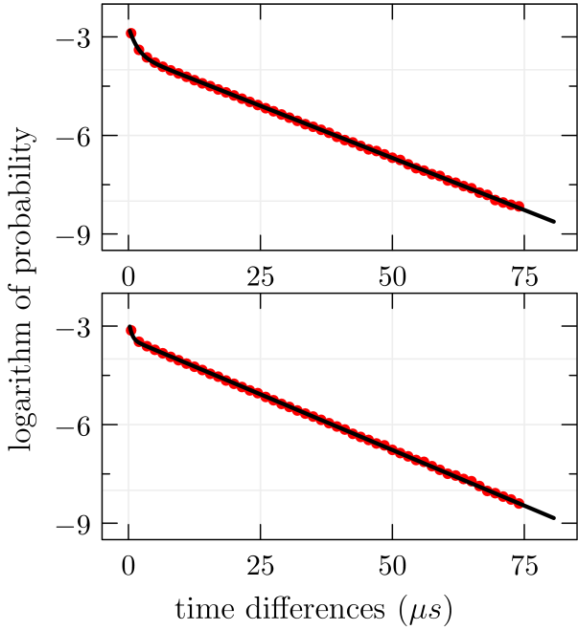

Fig. 3. Histograms of the detection probabilities binned in the time intervals for successive detections. By fitting the data, we acquire the desired P0 and τ for each detector.

A measurement of the HOM Visibility was then performed. The pulse width was set to 2 ns and the gate width at nominal width of 7 ns. Each of these widths can be changed by about 10% without appreciable change in the reported results. The dark counts were recorded for the two detectors at $10^{-4}$ and $4 \times 10^{-5}$ per gate, respectively while the dead time was set to 0.1 μs. Increasing the dead time further decreased the dark counts. In Fig. 4, the measurement results are presented in comparison to our model showing good agreement.

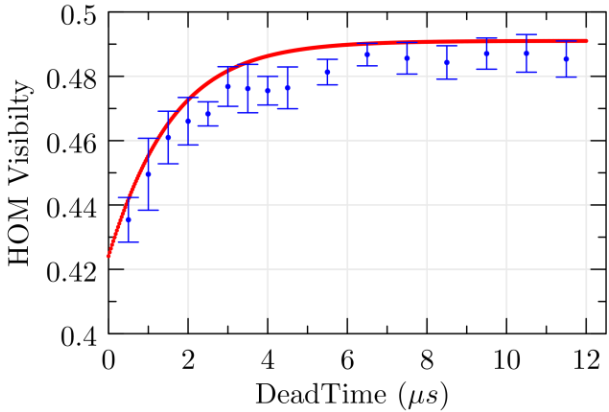

Fig. 4. HOM visibility *vs.* applied dead time at 2-MHz gate and pulse frequency.

For our second trial we used a gating and pulse frequency of 6 MHz, the photon number was fixed at 0.15 for both input arms.

The gate width was set at nominal value of 7 ns and pulse width at 2 ns.

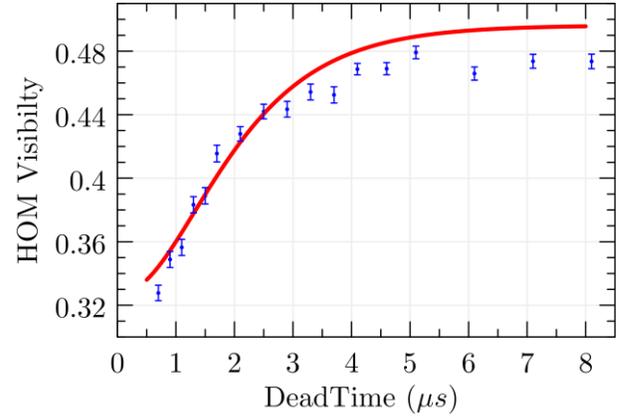

Fig. 5. HOM *vs.* dead-time data plotted with the theoretical curve (calculated from (17), (19), (20)). The pulse and triggering frequency was 6 MHz with a mean photon number of 0.15. Gate width was 7 ns, pulse width 2 ns.

### B. Source Effects on HOM Visibility

By lowering the total input intensity, the HOM visibility is improved. However, reaching very low intensities may render the experiment vulnerable to dark counts, and increases the required time to perform a measurement. This in turn renders the experiment vulnerable to various drifts (e.g., the drift in the state of polarization, or in the DC offset of the modulators).

We examined the effect of the overall input intensity on the HOM visibility. Setting the intensities of the input beams equal, $\mu_a = \mu_b = \mu$ in (11), (14), and (15), we studied the dependence of the HOM visibility on the average input photon number $\mu$. Theoretically, the HOM visibility approaches the limit value 0 at large input intensities, whereas it approaches the maximum value 0.5 at weak intensities.

In our measurements, we used 2-ns width pulses. Our detectors were running in external gating mode at 1-MHz trigger frequency with an effective gate width of approximately 3 ns (nominal gate width set to 7 ns) and 10% efficiency. The dark counts of the two detectors were recorded approximately as $2.5 \times 10^{-5}$ and $1.5 \times 10^{-5}$ per gate, respectively. The dead time on the detectors was set to 7 μs (a longer dead time does not change the results appreciably). To make sure that the beam-splitter inputs were equal, the free-space path of one arm was blocked, and the intensity of the unblocked armed was digitally attenuated until the detection rate reached the desired value. The average photon-number input to the beam splitter is related to the observed detection rate by

$$\mu = \frac{2}{\eta} \ln \left( \frac{1 - R_{det} T_{dt} + R_{det} T_{gat}}{1 - R_{det} T_{det}} \right), \quad (21)$$

where $\eta$ is the detector efficiency, $R_{det}$ is the detector rate of each unblocked input, $T_{dt}$ is the dead time, and $T_{gat}$ is the gating period. The factor of 2 in (21) accommodates the intensity splitting at the 50:50 beam splitter.

In Fig. (6), the measurement results of the HOM visibility as a function of the input photon number are presented and compared with the theoretical model (calculated using (11), (14), and (15)), showing good agreement between theory and experiment.

Next, we consider the effect of imperfections in input state preparation on the HOM visibility.

In a realistic experimental setup, two independent laser beams are independently attenuated. In practice, perfect intensity balance may be not possible. Using (11), (14), and (15), we can model the HOM visibility theoretically as a function of the ratio of the input photon numbers $\mu_a / \mu_b$.

For our measurements, the dead time for each detector was set to 7 μs with efficiency 10%. Each free-space arm was blocked for either Alice/Bob between data points to record detector count-rates. The count rates were controlled via digital attenuation and set to desired values to within 2%. From the detector count-rates and using formula (21), the photon number can be extracted for each count-rate. In this measurement the photon number for the input arm at port $\alpha$ was fixed at $\mu_a = 0.47$, while varying the attenuation on the input at port $b$ digitally. We sent weak coherent pulses at 1 MHz with pulse widths of 2 ns through the beam splitter and to our detectors. Each detector's gate width was approximately 7 ns to mitigate the detection of background source photons outside the intended pulse width. In Fig. 7, the measurement results of the HOM visibility as a function of the ratio of input photon numbers are plotted with the theoretical model (using (11), (14), and (15)), showing good agreement.

Next, we consider the effect of the polarization misalignment of the incoming beams on the HOM visibility. Equations (11), (14), and (15) show the dependence of the HOM visibility on the polarization misalignment $\Phi$ in (13). Assuming that the bases of the two inputs are perfectly aligned, we can write the polarization vectors $\hat{\varepsilon}_a$ and $\hat{\varepsilon}_b$ in terms of the transverse-electric (TE) and transverse-magnetic (TM) modes of the phase modulator's waveguide as:

$$\hat{\varepsilon}_a = \cos\phi_a \, |TE\rangle + \sin\phi_a e^{i\phi_0} \, |TM\rangle,$$
$$\hat{\varepsilon}_b = \cos\phi_b \, |TE\rangle + \sin\phi_b e^{i\phi_M} \, |TM\rangle, \tag{22}$$

where $\phi_M = (V_g/V_\pi) \cdot \pi$ is the modulation phase caused by the driving generator, $V_g$ is the voltage applied by the generator, and $V_\pi$ is the constant voltage that causes a $\pi$ phase shift. Using a manual polarization controller, we carefully arrange the input to the waveguide to be at $45°$ with respect to the waveguide's axis, so that $\cos\phi_a = \cos\phi_b = 1/\sqrt{2}$. The polarization misalignment angle $\Phi$ in (13) can then be related to the applied voltage as:

$$\cos\Phi = \left| \hat{\varepsilon}_a \cdot \hat{\varepsilon}_b^* \right| = \cos\frac{\pi V_g}{2V_\pi}. \tag{23}$$

For our measurements, we controlled the state of polarization using the Polarization Modulation setup described in [28]. We used a Keysight waveform generator to drive an EOSpace Phase Modulator. The $V_\pi$ voltage of the phase modulator was determined to be 5.25 V. Pulses of width 2 ns and average photon number $\mu = 0.45 \pm 0.05$ interfered at a 50:50 beam splitter. The outputs were directed to two SPADs operated at free-gated mode at 10% efficiency with gate period 1 μs, dead time set at 7 μs, and nominal gate width of 7 ns. The dark counts were recorded for the two detectors, approximately $5.5 \times 10^{-5}$ and $2.0 \times 10^{-5}$ per gate, respectively. The coincidence window was set at 5 ns.

Fig. 8 depicts the measured HOM visibility as a function of the polarization angle mismatch. Pulses of width 2 ns and frequency 1 MHz interfered at a 50:50 beam splitter. The relative polarization angle was modulated by a phase modulator with $V_\pi = 5.25$ V. SPADs of 10% efficiency and 7-μs dead time, 7-ns gate width operated at 1-MHz frequency. Experimental data were compared with the theoretical curve (calculated using (11), (14), and (15)) and good agreement was obtained.

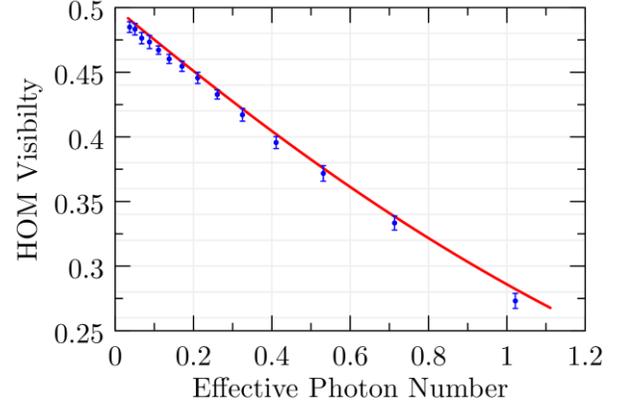

Fig. 6. HOM visibility vs effective photon number (ημ) in each input state with the two inputs kept at equal intensities. Pulses of 2-ns width interfere at the 50:50 beam-splitter. Outputs were directed to the two SPADs externally triggered at 1 MHz with a 7-μs dead time. Theoretical curve is calculated using (11), (14), and (15).

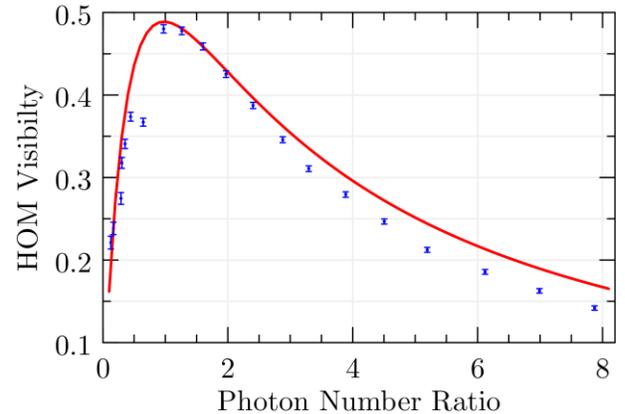

Fig. 7. HOM visibility *vs.* photon number ratio of inputs *a* and *b*. The photon number for *a* was fixed at 0.47 while the photon number for input *b* was varied via digital attenuation. Each detector's gating window was 7 ns. Theoretical curve calculated using (11), (14), and (15).

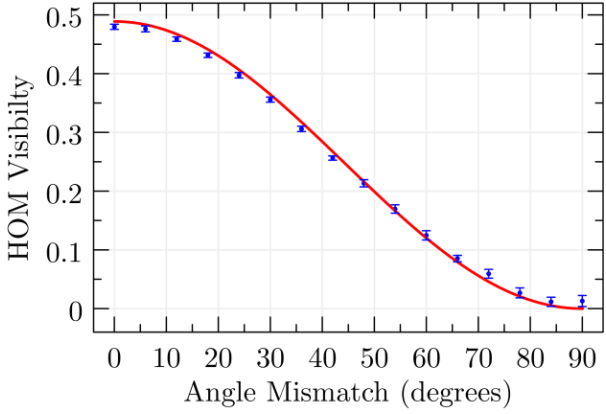

**Fig. 8.** HOM visibility *vs.* relative polarization angle. Pulses of width 2 ns and frequency 1 MHz interfere at the 50:50 beam splitter. Relative polarization angle is modulated with a phase modulator having $V_\pi = 5.25$ V. A dead time of 7 μs was used with a 7-ns gating window triggered at 1 MHz. Theoretical curve calculated using (11), (14), and (15).

## V. CONCLUSION

In this work, we parametrized the Hong-Ou-Mandel interference visibility in terms of realistic imperfections that may appear in experimental implementations using weak coherent states. We examined the effect of mismatches in the state of polarization and intensities of the inputs. We also considered imperfections on the detector side resulting from the detector's after pulses as well as the effect of the overall intensity of otherwise perfect sources. We conducted measurements that resulted in experimental data that agreed very well with our theoretical models.

In conclusion, good Hong-Ou-Mandel interference visibility is attainable using standard commercially available optical components and single-photon detectors. We conclude that the after-pulse effect can be effectively mitigated by applying a dead time $6 - 8$ μs, when the detectors are triggered at a few-MHz frequencies. Realistic intensity imbalances were less than 10% and they have minimal impact on the measured HOM visibility. For example, the HOM visibility is expected to be 0.489 for $\mu_a = \mu_b = 0.45$. A realistic imbalance $\mu_a = 0.45$ and $\mu_b = 0.50$ would decrease the visibility to just 0.487. Some extra care should be taken when adjusting the state of polarization for the two arms. Assuming $\mu_a = \mu_b = 0.45$, while a 0° misalignment gives a visibility of 0.489, a 6° misalignment gives a visibility of 0.483. Given that manual polarization controllers can achieve typical extinction ratios $20 - 30$ dB [29], a misalignment of that order should be expected. We identify the state of polarization misalignment as the major source of error in the measurements we present in this work. We finally discussed how the HOM visibility is affected by the overall input intensity aiming to achieve efficient intensities for practical measurements while remaining in the quantum regime.

## APPENDIX

Here we provide details of the theoretical model for the HOM visibility. We assume that the two detectors have efficiencies $\eta_{c,d}$ and dark count probabilities $d_{c,d}$, respectively, and are blind to the photon number, i.e., a single-photon event cannot be distinguished from a multi-photon event.

Let $P_{mn}^{(out)}$ be the probability that $m$ $(n)$ photons arrive at the detector at port $c$ $(d)$.

Since the ports $c$ and $d$ are separate, the output state (5) can be factorized into coherent states:

$$| \Psi_{out} \rangle = | \gamma_H \rangle \otimes | \gamma_V \rangle \otimes | \delta_H \rangle \otimes | \delta_V \rangle, \tag{A1}$$

where the coherent states with parameter $\gamma_i$ $(\delta_i)$ are in output port $c$ $(d)$, $i=H,V$, and

$$\gamma_i = (\alpha t \hat{\varepsilon}_a + \beta r \hat{\varepsilon}_b) \cdot \hat{\varepsilon}_i , \quad \delta_i = (\alpha r \hat{\varepsilon}_a - \beta t \hat{\varepsilon}_b) \cdot \hat{\varepsilon}_i . \tag{A2}$$

Therefore, we can write the probability $P_{mn}^{(out)}$ as a product:

$$P_{mn}^{(out)} = P_m^{(out,c)} P_n^{(out,d)}, \tag{A3}$$

where

$$P_m^{(out,c)} = \sum_{m_H + m_V = m} P_{m_H} P_{m_V},$$

$$P_m^{(out,d)} = \sum_{n_H + n_V = n} P_{n_H} P_{n_V}. \tag{A4}$$

The probabilities on the right-hand side are easily deduced from the corresponding coherent states. We obtain

$$P_{m_i} = e^{-|\gamma_i|^2} \frac{|\gamma_i|^{2m_i}}{m_i!}, \quad P_{n_i} = e^{-|\delta_i|^2} \frac{|\delta_i|^{2n_i}}{n_i!}. \tag{A5}$$

Using the binomial theorem, we deduce

$$P_m^{(out,c)} = e^{-\mu_c} \frac{\mu_c^m}{m!}, \quad P_n^{(out,d)} = e^{-\mu_d} \frac{\mu_d^n}{n!}, \tag{A6}$$

and therefore

$$P_{mn}^{(out)} = e^{-\mu_c - \mu_d} \frac{\mu_c^m \mu_d^n}{m! n!}, \tag{A7}$$

where

$$\begin{aligned} \mu_c &= \sum_{i=H,V} |\gamma_i|^2 = | \alpha t \hat{\varepsilon}_a + \beta r \hat{\varepsilon}_b |^2 \\ &= | \alpha |^2 t^2 + | \beta |^2 r^2 + 2 | \alpha \beta | tr \cos \Phi \cos(\theta_a - \theta_b + \phi_0), \\ \mu_d &= \sum_{i=H,V} |\delta_i|^2 = | \alpha r \hat{\varepsilon}_a - \beta t \hat{\varepsilon}_b |^2 \\ &= | \alpha |^2 r^2 + | \beta |^2 t^2 - 2 | \alpha \beta | tr \cos \Phi \cos(\theta_a - \theta_b + \phi_0). \end{aligned} \tag{A8}$$

Notice that $\phi_o$ is an irrelevant phase, because we average over the phases.

The probability of detection if $m(n)$ photons reach detector $c$ $(d)$ is $1 - (1 - \eta_c)^m (1 - d_c)(1 - (1 - \eta_c)^n (1 - d_d))$. Therefore, the probability of detection given $m$ $(n)$ photons coming out of beam splitter port $c$ $(d)$ is

$$\begin{aligned} P_{mn} &= \Big[ 1 - (1 - \eta_c)^m (1 - d_c) \Big] \\ &\times \Big[ 1 - (1 - \eta_d)^n (1 - d_d) \Big] P_{mn}^{(out)}. \end{aligned} \tag{A9}$$

The total coincidence probability is

$$P^{(coin)} = \sum_{m,n=0}^{\infty} P_{mn} = \Big( 1 - e^{-\eta_c \mu_c} (1 - d_c) \Big) \Big( 1 - e^{-\eta_d \mu_d} (1 - d_d) \Big) \tag{A10}$$

showing that the effective average photon number is the average photon number of the output beam that reaches the detector multiplied by the detector efficiency.

After averaging over the phases $\theta_{a,b}$, we obtain an expression in terms of Bessel functions,

$$
\begin{aligned}
P^{(coin)} \to \int_0^{2\pi} \frac{d\theta_a}{2\pi} \int_0^{2\pi} \frac{d\theta_b}{2\pi} P^{(coin)} \\
= 1 - \mathcal{C}I_0(2\eta_c \sqrt{\mu_a \mu_b}\, tr \cos\Phi) \\
- \mathcal{D}I_0(2\eta_d \sqrt{\mu_a \mu_b}\, tr \cos\Phi) \\
+ \mathcal{C}\mathcal{D}I_0(2(\eta_c - \eta_d)\sqrt{\mu_a \mu_b}\, tr \cos\Phi),
\end{aligned}
\tag{A11}
$$

where

$$
\begin{aligned}
\mathcal{C} = e^{-\eta_c(\mu_a t^2 + \mu_b r^2)}(1 - d_c), \\
\mathcal{D} = e^{-\eta_d(\mu_a r^2 + \mu_b t^2)}(1 - d_d).
\end{aligned}
\tag{A12}
$$

For the HOM visibility, we also need to calculate the probabilities for one of the two detectors to click. The probability for detector at port $c$ to click, after averaging phases, is

$$
P^{(c)} = 1 - \mathcal{C}I_0(2\eta_c \sqrt{\mu_a \mu_b}\, tr \cos\Phi).
\tag{A13}
$$

Similarly, for the other detector, we obtain

$$
P^{(d)} = 1 - \mathcal{D}I_0(2\eta_d \sqrt{\mu_a \mu_b}\, tr \cos\Phi).
\tag{A14}
$$

We define the HOM visibility by

$$
V_{HOM} = 1 - \frac{P^{(coin)}}{P^{(c)}P^{(d)}}.
\tag{A15}
$$

Notice that $V_{HOM} = 0$, for $\Phi = \pi/2$ (orthogonal polarizations).

In the limit $\alpha,\ \beta \to 0$ (small average photon number), and in the ideal case of no dark counts ($d_c = d_d = 0$), the HOM visibility is approximately

$$
V_{HOM} \approx \frac{2tr\,\mu_a \mu_b \cos^2\Phi}{(t\mu_a + r\mu_b)(r\mu_a + t\mu_b)}.
\tag{A16}
$$

Its maximum value of $tr$ is attained for $\Phi = 0$, and $\mu_b/\mu_a = t/r$.

For a 50:50 beam splitter, it reduces to

$$
V_{HOM} \approx \frac{2\mu_a \mu_b \cos^2\Phi}{(\mu_a + \mu_b)^2},
\tag{A17}
$$

which vanishes for $\Phi = \pi/2$ (orthogonal polarizations), and for $\Phi = 0$ (parallel polarizations), it has maximum $1/2$ at $\mu_a = \mu_b$.

ACKNOWLEDGMENT

We thank Daniel Gauthier for useful comments.